\documentclass[pdflatex,sn-nature]{sn-jnl}

\usepackage{graphicx}%
\usepackage{multirow}%
\usepackage{amsmath,amssymb,amsfonts}%
\usepackage{amsthm}%
\usepackage{mathrsfs}%
\usepackage[title]{appendix}%
\usepackage{xcolor}%
\usepackage{textcomp}%
\usepackage{manyfoot}%
\usepackage{booktabs}%
\usepackage{algorithm}%
\usepackage{algorithmicx}%
\usepackage{algpseudocode}%
\usepackage{listings}%
\usepackage{soul}
\usepackage{siunitx}

\theoremstyle{thmstyleone}%
\theoremstyle{thmstyletwo}%

\theoremstyle{thmstylethree}%

\begin{document}

\title[Article Title]{All-optical switching of nonlinear structured light in crystal-engineered van der Waals materials}



\author[1]{\fnm{Paolo} \sur{Valisa}}

\author[1,2]{\fnm{Marc} \sur{Richstaetter}}

\author[1]{\fnm{Bianca} \sur{Sanfilippo}}

\author[1]{\fnm{Benedikt} \sur{Ursprung}}

\author[2]{\fnm{Zhi Hao} \sur{Peng}}

\author[2,3]{\fnm{Victoria} \sur{Quirós-Cordero}}

\author[1]{\fnm{Francesco} \sur{Gucci}}

\author[3]{\fnm{Xiaoyang} \sur{Zhu}}

\author[2]{\fnm{P. James} \sur{Schuck}}

\author[1]{\fnm{Giulio} \sur{Cerullo}}

\author*[4]{\fnm{Luca} \sur{Carletti}}\email{luca.carletti@unibs.it}

\author*[1]{\fnm{Chiara} \sur{Trovatello}}\email{chiara.trovatello@polimi.it}

\affil[1]{\orgdiv{Dipartimento di Fisica}, \orgname{Politecnico di Milano}, \orgaddress{\street{Piazza Leonardo da Vinci, 32}, \city{Milano}, \postcode{20133}, \country{Italy}}}

\affil[2]{\orgdiv{Department of Mechanical Engineering}, \orgname{Columbia University}, \orgaddress{\city{New York}, \postcode{10027}, \state{New York}, \country{United States}}}

\affil[3]{\orgdiv{Department of Chemistry}, \orgname{Columbia University}, \orgaddress{\city{New York}, \postcode{10027}, \state{New York}, \country{United States}}}

\affil[4]{\orgdiv{Dipartimento di Ingegneria dell'Informazione}, \orgname{Università degli Studi di Brescia}, \orgaddress{\street{Via Branze 38}, \city{Brescia}, \postcode{25123}, \country{Italy}}}

\abstract{The orbital angular momentum (OAM) of light is a discrete, unbounded degree of freedom that underpins mode-multiplexed communications and high-dimensional quantum photonics. Yet, dynamic OAM control remains dependent on bulky free-space optics or cascaded architectures that separate switching from wavefront shaping, hindering nanoscale integration. Here, we engineer artificial van der Waals crystals from rhombohedrally stacked (3R) MoS$_2$, in which spatial control of the local crystal orientation imprints a nonlinear geometric phase onto the second-harmonic (SH) field, enabling background-free generation of SH vortex beams in an ultrathin  (46 nm) van der Waals platform. Leveraging the C$_{3v}$ symmetry of 3R-MoS$_2$, we demonstrate monolithic, all-optical switching with sub-optical-cycle precision between Hermite-Gauss-like and Laguerre-Gaussian vortex SH beams with opposite topological charges ($l=\pm1$). Our results establish artificial 3R-MoS$_2$ crystals as a monolithic platform for the generation and all-optical reconfiguration of nonlinear structured light at the nanoscale, advancing active nanophotonic sources for integrated classical and quantum photonic technologies.}


\maketitle

The ability to structure light by controlling its amplitude, phase, polarization, and propagation direction has become increasingly important in modern photonics\cite{Forbes2021,Rubinsztein-Dunlop_2017,FrankeArnold2008}. Among the available degrees of freedom, the orbital angular momentum (OAM) of light has emerged as a key resource for both classical and quantum technologies\cite{Alen92}. OAM-carrying beams exhibit a helical wavefront with an azimuthal phase dependence $(e^{i\ell\phi})$, where the integer ($\ell$) is the topological charge and corresponds to an OAM of ($\ell\hbar$) per photon. Unlike spin angular momentum, which corresponds to two orthogonal polarization states, OAM can in principle span an unbounded set of mutually orthogonal spatial modes. This high-dimensional state space is thus central in mode-division-multiplexed optical communications \cite{Trichili2019Communication,Willner2015Communication,Wang2012Communication,Bozinovic2013Communication}, quantum information processing \cite{Groblacher2006Quantum,Berkhout2010Quantum,Fickler2012Quantum,Mirhosseini2015Quantum,Malik2016Quantum,Cozzolino2019Quantum}, structured-light microscopy \cite{Hell1994Imaging,Furhapter2005aImaging,Furhapter2005bImaging,Steiger2012Imaging}, and optical manipulation \cite{Paterson2001Manipulation,GarcesChavez2003Manipulation}. 

Realizing this potential in the active and monolithically integrated chip-scale architectures envisioned for next-generation photonic circuits requires microscopic, chip-integrable and dynamically reconfigurable OAM sources\cite{Forbes2024}. To date, OAM control has almost exclusively been imposed downstream of the light source, using spatial light modulators\cite{Carpentier2008}, cylindrical lenses\cite{Beijersbergen1993}, or spin-to-orbital angular momentum conversion in anisotropic media\cite{Marrucci2006} or tightly focused beams\cite{Zhao2007}. However, these technological solutions are generally bulky, static or limited by relatively long reconfiguration times, typically in the microsecond-to-millisecond range.

Optical metasurfaces have provided a compact alternative for generating vortex beams carrying OAM, exploiting, \textit{e.g.}, the geometric or Pancharatman-Berry (PB) phase, where the phase of the transmitted of reflected radiation is controlled through the spatially varying orientation of anisotropic nanostructures\cite{Genevet2012,Gorodetski2013,Genevet2015,Guo2020, DeOliveira2023,MetaRoadmap2024}. Extending this approach to nonlinear optics enables structured light to be generated directly at new frequencies and virtually background-free, while introducing additional degrees of freedom associated with the nonlinear susceptibility and light-matter interaction\cite{Buono2022,PinheiroDaSilva2022,Tang2020,deCeglia2024}. Recently, generation of OAM-carrying vortex beams from nonlinear metasurfaces has been demonstrated at the second\cite{Guercio2026,KerenZur2016,Rong2023,Coudrat2025} and third harmonics\cite{Wang2018,ReinekeMatsudo2022}, establishing a route toward compact sources of structured light.

In this framework, layered van der Waals materials, \textit{e.g.}, semiconducting transition metal dichalcogenides (TMDs), are particularly attractive for nonlinear light generation at ultracompact spatial scales, due to the large optical nonlinearity ($\chi^{(2)}\sim100-\SI{1000}{pm/V}$) and reduced dimensionality\cite{Trovatello2024Perspective,GarciadeAbajo2025Roadmap}. Moreover, due to the strong in-plane covalent bonds and weak out-of-plane van der Waals forces, TMDs can be exfoliated into atomically thin flakes and readily integrated onto virtually any substrate due to relaxed lattice matching requirements\cite{FangLiu2020,Zhang2026,Wang2026,Lin2026}. While TMDs have spatially isotropic in-plane linear dielectric response, their second-order nonlinear susceptibility tensor $\chi^{(2)}$ is strongly anisotropic owing to the crystal's point-group symmetry\cite{Boyd,Cotter}. 

This symmetry-enabled anisotropy has already been exploited to demonstrate close to unity-depth, femtosecond-limited all-optical modulation of the linear polarization and amplitude of SHG in monolayer MoS$_2$\cite{Klimmer2021}. It has also been used to coherently control the circular polarization state and chirality of harmonic generation in monolayer MoS$_2$ by interfering counter-rotating circularly polarized pump pulses\cite{ZhangSun2022chiral}, also for the implementation of chiral logic gates\cite{Zhang2022chiralgates}. These results establish that the crystal symmetry of TMDs can be harnessed to switch polarization and chirality of nonlinear signals on ultrafast timescales using light. More recently, nonlinear frequency mixing of optical vortices in monolayer TMDs has demonstrated the transfer and conversion of OAM through difference- and sum-frequency generation and four-wave mixing, enabling control over the topological charge and radial mode of the generated fields\cite{Norden2025}. In these experiments, however, the OAM of the nonlinear output was determined by that of the incident fields through OAM conservation, rather than actively controlled through the crystal-symmetry-dependent nonlinear response. Whether crystal symmetry can instead be harnessed to all-optically switch the OAM of a nonlinear beam therefore remains an open question.

Initial steps towards nonlinear wavefront engineering in TMDs have relied on spatial patterning of either the nonlinear material\cite{Lochner2019,Dasgupta2019,Dasgupta2020} 
or its surrounding photonic environment\cite{Hu2019,Busschaert2019,Spreyer2020,Hong2020
,Zhao2021}. A related theoretical proposal has shown that position and orientation of TMD meta-atoms can be engineered to impart nonlinear geometric phases and generate structured SH fields with tailored polarization and OAM\cite{Meng2020Theory}. Beyond TMDs, new layered ferroelectrics such as niobium oxyhalides (NbOX$_2$) have also been exploited to generate fork holograms \cite{Deka2025}. These studies have established van der Waals layered materials as prominsing ultrathin platforms for nonlinear holography and wavefront shaping. However, the encoded phase profiles were defined lithographically and therefore remained static after fabrication.

More recently, dynamic wavefront control was introduced by cascading the symmetry-enabled polarization switching of SHG in a monolayer WSe$_2$\cite{Klimmer2021} with a separate, helicity-sensitive dielectric metasurface\cite{Sinelnik2024}. By varying the temporal delay between two orthogonally polarized femtosecond pump pulses, the polarization and, after a quarter-wave plate, the helicity of the generated SH could be switched on a pulsewidth-limited timescale\cite{Sinelnik2024}.
While advancing ultrafast structured-light modulation, this demonstration still relied on a cascaded architecture in which frequency conversion and wavefront encoding occurred in physically distinct components. Furthermore, nonlinear frequency conversion in monolayer TMDs has always been limited by the poor conversion efficiencies ($\eta$) due to their sub-nm thickness $z$ ($\eta\propto z^2$)\cite{Trovatello2021}. 

Unlike the centrosymmetric 2H polytype, whose even-layer stacks are SHG-forbidden in the electric-dipole approximation, the rhombohedrally stacked (3R) TMDs belong to the C$_{3v}$ point group and possess broken inversion symmetry for any number of layers, enabling quadratic thickness-dependent efficiency scaling\cite{Zhao2016,Xu3R2022}. This coherent buildup can be further extended to quasi-phase-matching schemes\cite{Trovatello2025,Braun2026} as well as nonlinear metasurfaces\cite{Zograf2024,Peng2025,Zhu2026} to achieve macroscopic SH conversion efficiencies, also with the possibility to realize all-optical SH polarization switching with near-unity modulation depth\cite{Seidt2025}. 3R-TMDs thus combine efficiency scaling with thickness with the tensor symmetry nonlinear anisotropy that enables ultrafast polarization and chirality control. This makes 3R-MoS$_2$ a suitable platform for extending this control to the OAM of the generated SH.

Here we design and realize artificial van der Waals crystals, based on 3R-MoS$_2$, to engineer the nonlinear geometric phase through spatial control of the crystal orientation. This enables the direct generation of SH vortex beams in layered compounds as thin as 46 nm. Furthermore, within a single crystal-engineered 3R-MoS$_2$, we demonstrate all-optical switching between vortex beams with either $l=-1$ or $l=+1$ topological charge and Hermite-Gaussian (HG) beams (with net zero topological charge), by controlling the delay of two fundamental frequency (FF) pulses in a collinear geometry with sub-optical-cycle precision. Our work establishes crystal-engineered van der Waals materials as a monolithic, ultracompact platform for the generation and dynamic reconfiguration of nonlinear structured light, collapsing frequency conversion and wavefront shaping into a single atomically-thin medium and opening a route towards on-chip, all-optically reconfigurable sources of classical and quantum structured light.

%
\begin{figure}[H]
    \centering
    \includegraphics[width=\linewidth]{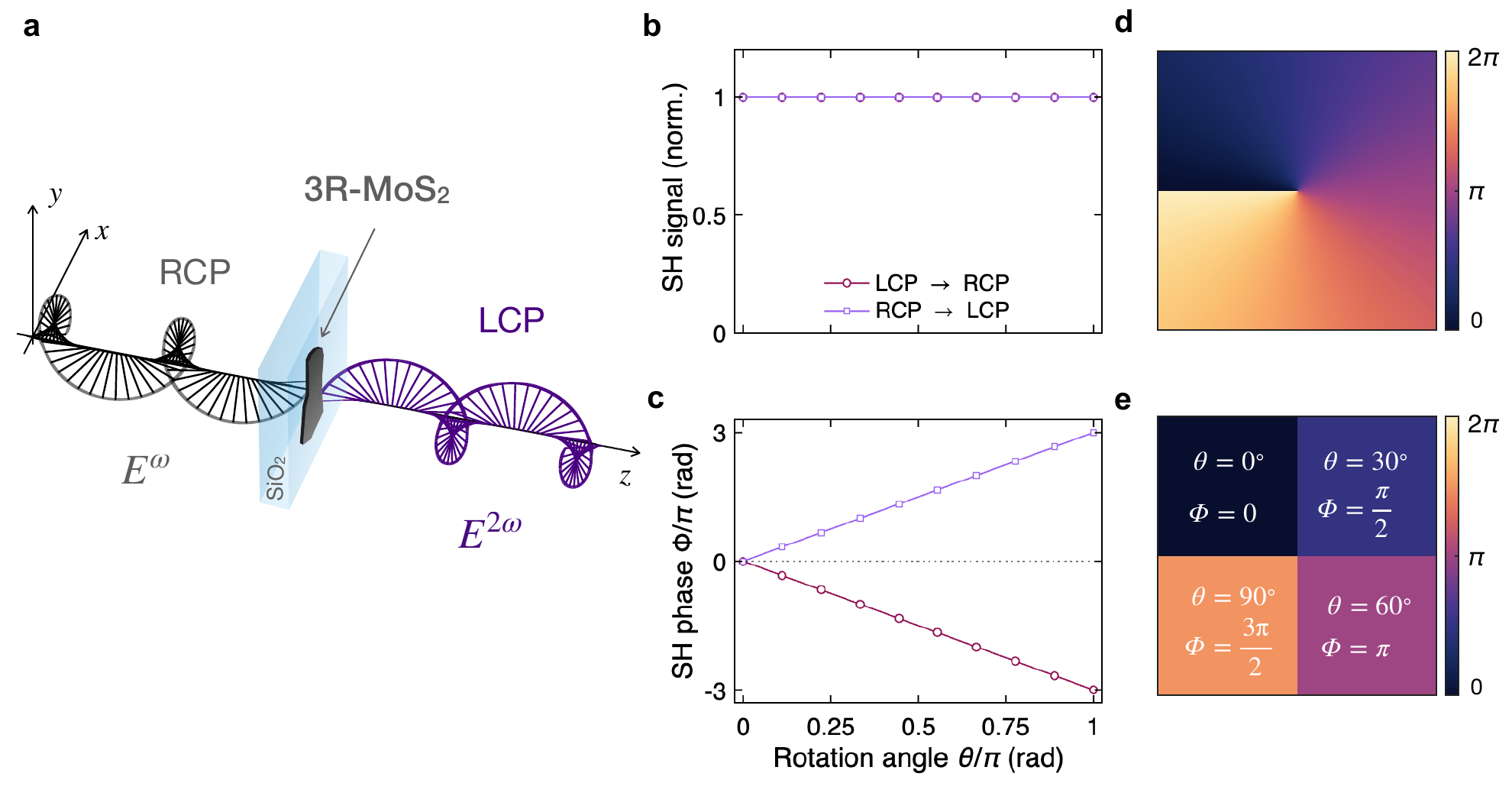}
    \caption{\textbf{Tensor-driven geometric phase in 3R-MoS\(_2\)}. \textbf{a} Schematics of the SHG in a bare 3R-MoS\(_2\) flake, for a circularly polarised (CP) FF. The emitted SH field has opposite helicity with respect to the FF due to TAM conservation. \textbf{b,c} Calculated intensity (b) and phase (c) of circularly polarized SH as a function of the in-plane rotation angle \(\theta\), \textit{i.e.}, of the crystal orientation with respect to the laboratory frame, for the two helicities of the FF. The legend indicates the pump polarization state $\rightarrow$ SH polarization state. \textbf{d} Spiral phase profile associated to the topological charge \(+1\).  \textbf{e} Four-level discretization of the spiral phase mask in (d). The plane is divided into four quadrants, with the crystal orientation \(\theta\) in each quadrant chosen according to the tensor-driven geometric-phase relation shown in (c).}
    \label{fig:Fig1}
\end{figure}
\noindent 3R-MoS\(_2\) belongs to the crystallographic point group \(C_{3v}\), which induces a second-order nonlinear susceptibility tensor whose nonzero elements are:
\begin{align}
    \chi^{(2)}_{yyy} = -\chi^{(2)}_{yxx} = -\chi^{(2)}_{xxy} = -\chi^{(2)}_{xyx} \equiv \chi^{(2)} .
    \label{eq:chi2}
\end{align}
\noindent From Eq.~(\ref{eq:chi2}), it is apparent that the nonlinear tensor transforms under in-plane rotations. In the case of a pristine flake, whose crystallographic axes are rotated by an angle \(\theta\) with respect to the laboratory frame, the \(x\) and \(y\) components of the induced nonlinear polarization vector in the SH, \( \mathbf{P^{2\omega}}=\varepsilon\bar{\mathbf{\chi^{(2)}}}\vdots \mathbf{EE} \), can be expressed as a function of the angle \(\theta\) as (see the Supplementary Note 1 for the derivation):
\begin{align}
    P_x^{2\omega}(\theta) = \varepsilon_0 \chi^2 ( \sin 3\theta (E_x^2 - E_y^2) - 2\cos 3\theta E_x E_y ) \label{eq:Px}\\  
    P_y^{2\omega}(\theta) = \varepsilon_0 \chi^2 ( -\cos 3\theta (E_x^2 - E_y^2) - 2\sin 3\theta E_x E_y ) \label{eq:Py} 
\end{align}
The expressions in Eqs.~(\ref{eq:Px}) and (\ref{eq:Py}) result in the typical six-lobe SH polarization-resolved intensity pattern, that has already been demonstrated for 3R-MoS\(_2\) \cite{Xu3R2022,Zograf2024}. The anisotropy of the second-order nonlinear susceptibility introduces a tensor-driven geometric phase in the SH response, as recently shown in AlGaAs metasurfaces with circularly polarized light\cite{Guercio2026}. 

When dealing with circular polarizations, the SH selection rules can be derived by total angular momentum (TAM) projection conservation principles\cite{Menshikov2025,Guercio2026}. If we define the TAM projections of FF and SH as \(m^\omega\) and \(m^{2\omega}\), respectively, we have that \(m^{2\omega}=2m^{\omega}+m_\chi\), where \(m_\chi\) is the additional momentum induced by the nonlinear response of the material. For 3R-MoS\(_2\), the three-fold rotational symmetry provides an additional momentum \(m_\chi=\pm3\). Thus, for a circularly polarized pump beam with a Gaussian field intensity profile (\(m^{\omega}=\pm1\)), we find that the allowed TAM projections for the SH with non-zero emission along the vertical directions are \(m^{2\omega}=-1\) for \(m^{\omega}=1\) and \(m^{2\omega}=1\) for \(m^{\omega}=-1\). This shows that, for circularly polarized FF, the generated SH will have opposite helicity with respect to the FF, as depicted in Fig.~\ref{fig:Fig1}a.

To derive the phase relations between the FF and SH polarization states and the rotation angle \(\theta\) of the 3R-MoS\(_2\) crystal, we operate a transformation from the Cartesian reference frame \((x,y,z)\) to the circular-polarization basis \((R,L,Z)\) (see Supplementary Note 1 for more details). The nonlinear polarization induced in the material by a pump field propagating along the positive \(z\)-axis direction for SH radiation emitted in the same direction can be written as (see the Supplementary Note 1 for derivation):
\begin{align}
    P_{R^+}^{2\omega}(\theta) &= -i\varepsilon_0\chi^{(2)}\sqrt{2}\left(E^\omega_{L^+}\right)^2 e^{-i3\theta},
    \label{eq:Pr}\\
    P_{L^+}^{2\omega}(\theta) &= i\varepsilon_0\chi^{(2)}\sqrt{2}\left(E^\omega_{R^+}\right)^2 e^{i3\theta}.
    \label{eq:Pl}
\end{align}
Equations~\eqref{eq:Pr} and~\eqref{eq:Pl} show that the forward-emitted SH field has opposite circular-polarization helicity with respect to the FF (as expected from TAM projection conservation) and acquires a geometric phase of \(\pm 3\theta\), with the sign determined by the FF helicity. The factor $3\theta$ is consistent with the threefold rotational symmetry, since rotating the crystal by $2\pi/3$ leaves the nonlinear response invariant while changing the geometric phase by $2\pi$. These results are verified by numerical electromagnetic calculations of the SH generated by a 100-\SI{}{nm}-thick 3R-MoS\(_2\) flake on a fused quartz substrate (see Supplementary Note 2 for more details). Although a specific flake thickness is used in the simulations, the tensor-induced geometric phase is independent of this parameter because it does not originate from propagation effects. Moreover, this mechanism is intrinsically broadband, as it does not rely on optical or material resonances. The calculated amplitude and phase of the generated SH field as a function of the FF helicity and the rotation angle \(\theta\) of the 3R-MoS\(_2\) crystal are shown in Fig.~\ref{fig:Fig1}b-c. 
It can be seen that, while the conversion efficiency remains constant as a function of \(\theta\), the SH generated by right-circularly polarised (RCP) FF acquires a phase factor of \(3\theta\) while the SH generated by left-circularly polarised (LCP) FF acquires a phase factor of \(-3\theta\).

Interestingly, the tensor-driven geometric phase appears also when the linearly polarized SH emission is considered. Indeed, we can express the nonlinear polarizations for linearly polarized SH from circularly polarized electric fields of the FF (see the Supplementary Note 1 for the derivation):
\begin{align}
P^{2\omega}_x(\theta)&=i\varepsilon_0\chi^{(2)}\left(e^{-i3\theta}E_{L^+}^2-e^{i3\theta}E_{R^+}^2\right)  \label{eq:Px_CPL} \\ 
P^{2\omega}_y(\theta)&=-\varepsilon_0\chi^{(2)}\left(e^{-i3\theta}E_{L^+}^2+e^{i3\theta}E_{R^+}^2\right)  \label{eq:Py_CPL} 
\end{align}
This shows that for a FF with a definite helicity, both linear polarization components of the SH inherit the same geometric phase factor \(e^{\pm i3\theta}\).

To experimentally validate the framework, we design a spiral phase mask (Fig. \ref{fig:Fig1}d) discretized as depicted in Fig.~\ref{fig:Fig1}e for the generation of structured light carrying principal topological charge equal to \(\pm 1\). The mask is composed of four quadrants where the crystal axes are oriented at four different angles \(\theta\) such that they represent a sampling of a total phase shift of \(2\pi\)~rad (Fig.~\ref{fig:Fig1}e). 

\begin{figure}[H]
    \centering
    \includegraphics[width=1\linewidth]{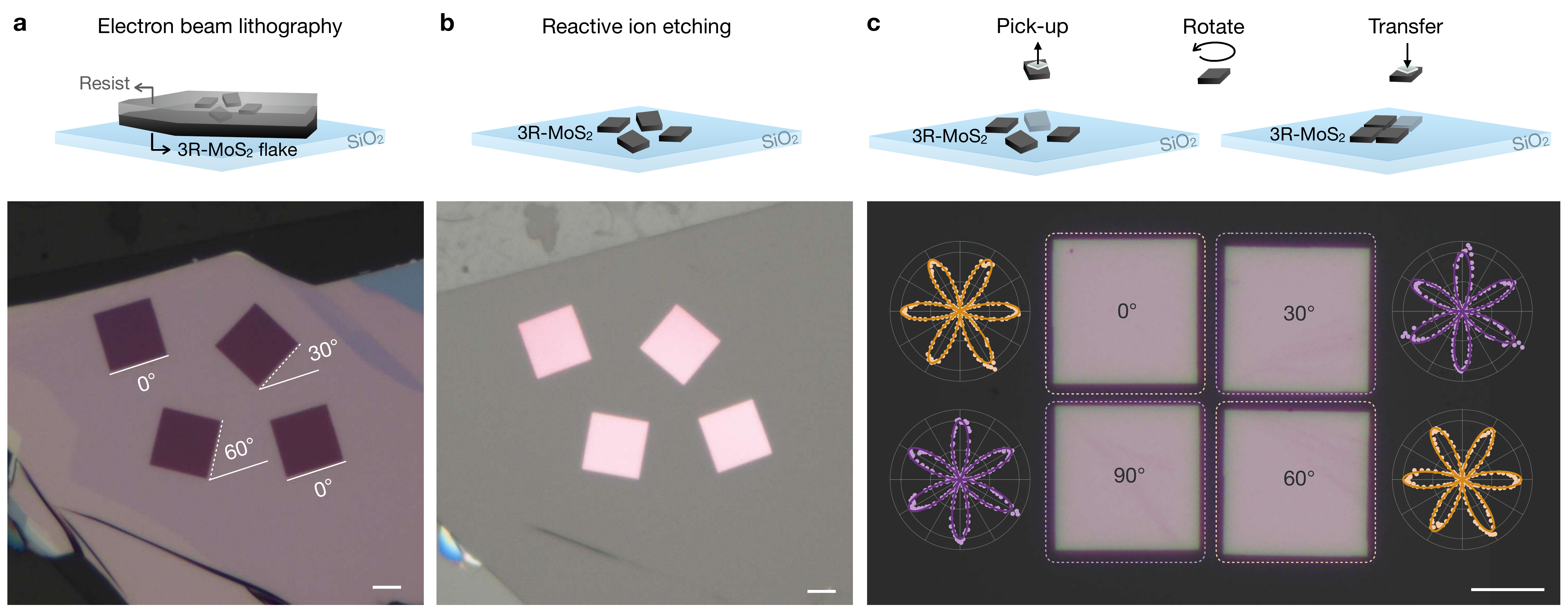}
    \caption{\textbf{Fabrication process of the crystal-engineered 3R-MoS$_2$ sample and SHG characterisation}. \textbf{a} Micrograph of the 3R-MoS$_2$ flake after electron beam lithography exposure for patterning. The four squared regions are marked with their relative crystal orientation. Flake thickness \SI{46}{nm}. \textbf{b} Isolated 3R-MoS$_2$ slabs via reactive ion etching. \textbf{c} Micrograph of the final disposition of the patterned 3R-MoS$_2$ squared slabs with identical thickness (\SI{46}{nm}) after pick up and transfer. Inset: polarisation-resolved SHG measurements of the four individual squares. Scale bars $\SI{10}{\micro m}$.}  
    \label{fig:Fig2}
\end{figure}

\noindent To realize the designed van der Waals artificial crystal structures, we fabricate individual flakes using standard dry mechanical exfoliation from commercial 3R-MoS$_2$ bulk crystals (HQ Graphene). Flakes with suitably large lateral size ($\SI{100}{\micro m}\times\SI{100}{\micro m}$) are then selected for patterning and transfer. Being that the nonlinear geometric phase encoding is independent of the flake thickness (see Supplementary Note 7), all the data shown in this work are measured on a crystal-engineered MoS$_2$ sample with thickness of $\SI{46}{nm}$. 

Next, we pattern the flake into four squares each with lateral size $\SI{20}{\micro m}\times\SI{20}{\micro m}$ and different crystal orientation, using electron beam lithography (see Fig. \ref{fig:Fig2}a). We then isolate the squared slabs using reactive ion etching, as shown in Fig. \ref{fig:Fig2}b. Finally, using polycarbonate polymer stamps we individually pick up each of the four square slabs, and adjacently stack them with relative crystal orientation of \SI{0}{\degree}, \SI{30}{\degree}, \SI{60}{\degree} and \SI{90}{\degree}, respectively, as shown in Fig. \ref{fig:Fig2}c (for further details on the sample fabrication see Supplementary Note 3). 

To confirm the $\SI{30}{\degree}$ offset between adjacent squares as by design, we measure the relative orientations of the four quadrants of the final structure via polarization-resolved SHG \cite{Malard2013} on each individual square slab, as shown by the flower pattern insets of Fig. \ref{fig:Fig2}c (for further details on the sample characterization see Supplementary Note 3).

\begin{figure}[H]
    \centering
    \includegraphics[width=.9\linewidth]{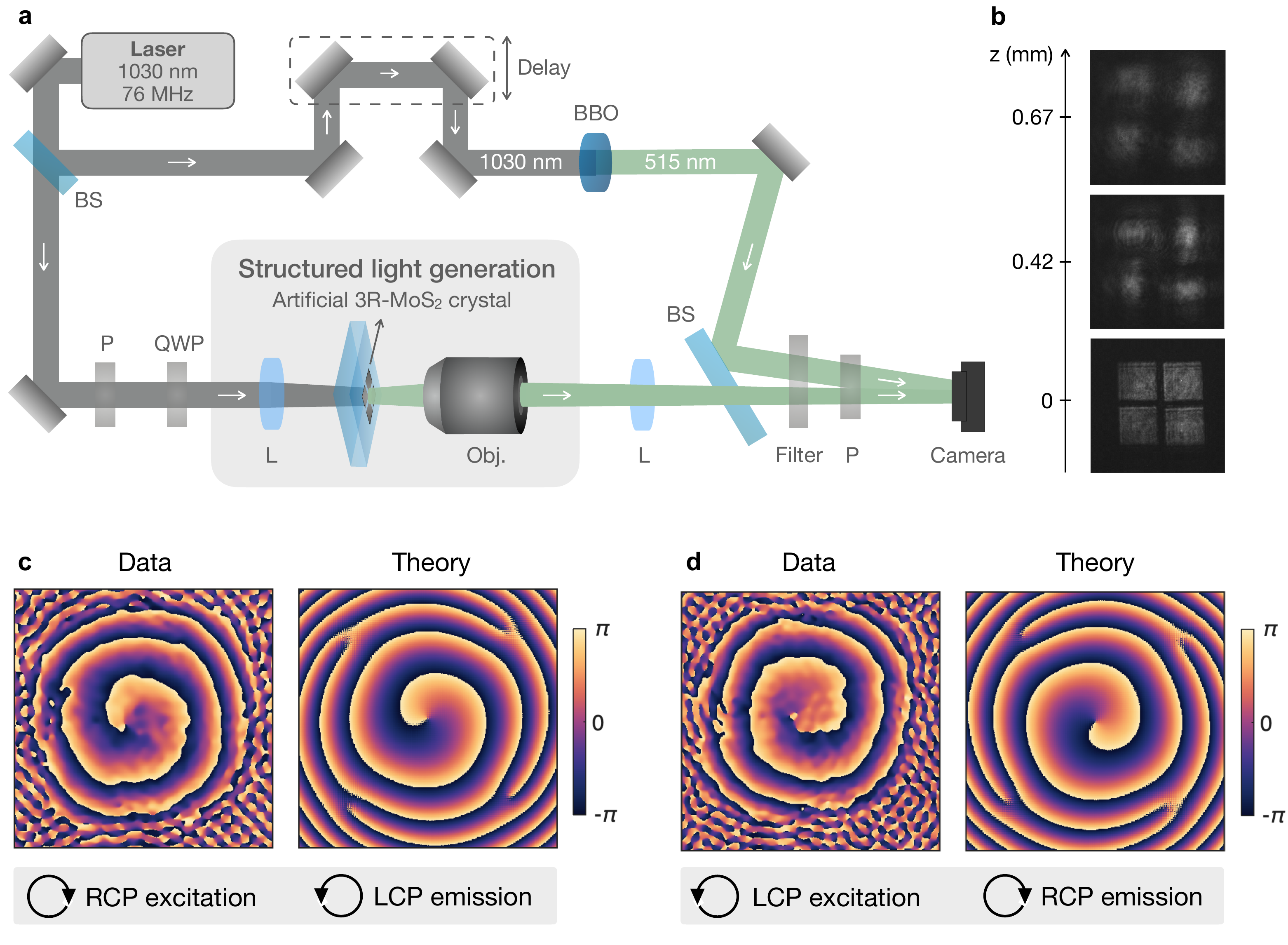}
    \caption{\textbf{Nonlinear structured light vortex generation in crystal-engineered 3R-MoS$_2$. a} Experimental setup: generalized Mach-Zehnder interferometer for the characterization of nonlinear structured light vortexes generated in crystal-engineered 3R-MoS$_2$. A pulsed laser source at 1030 nm is split into two beams: one to excite the $\SI{46}{nm}$-thick 3R-MoS$_2$ sample and generate SH light vortexes and one reference beam for off-axis digital holography. BS: beam splitter. P: polariser. QWP: quarter waveplate. L: lens. Obj: objective. \textbf{b} Images of SHG vortexes obtained with the objective's focal plane at a different distances from from the sample's plane. \textbf{c-d} Experimental and simulated phase of the vortex beam for RCP (c) and LCP (d) excitation.}  
    \label{fig:Fig3}
\end{figure}

\noindent A custom-made transmission microscope is used to generate nonlinear structured light with the realized 46-\SI{}{nm}-thick crystal-engineered 3R-MoS$_2$ sample. The experimental setup, depicted in Fig. \ref{fig:Fig3}a, features an off-axis digital holographic microscope for the detection of the SH light vortexes, seeded by a linearly polarised ytterbium-based pulsed laser source at 1030 nm (150 fs, 76 MHz). A beam splitter (BS) is used to split the 1030 nm light in an excitation beam - used to generate SH in the 3R-MoS$_2$ sample at $\SI{515}{nm}$ - and a  reference beam, also frequency doubled to $\SI{515}{nm}$ in a 2-mm-thick $\beta$-barium borate (BBO) crystal. The pump branch is sent through a quarter waveplate (QWP) for polarization control and focused with a 75 mm lens (L) on the sample. The focal length and position of the lens is chosen such that the FWHM of the pump beam profile matches the lateral dimensions of the sample ($\SI{40}{\micro m}\times\SI{40}{\micro m}$). The SH light emitted by the sample is then collected by a $25\times$ objective with numerical aperture (NA) of 0.5, spectrally filtered to remove the residual pump light and imaged onto a CMOS camera using a $\SI{200}{mm}$ lens. 

To reconstruct the spatial distribution of the emitted SH phase, we let the second (reference) branch recombine with the object arm on a BS combiner to realise an off-axis digital holography detection system\cite{Holography}, as depicted in Fig. \ref{fig:Fig3}a (see Methods). The temporal overlap of SH and reference beams is optimised using a micromechanical delay line placed on the reference beam path. A polariser after the combining BS is inserted to linearly polarise object and reference fields along the same axis, so as to maximise the visibility of interference fringes regardless of the phase and amplitude changes induced by the reflection and transmission through the combiner. This configuration is valid because the CP SH field preserves its phase also after the linear polariser, as further demonstrated via the implementation of complementary detection strategies (see Supplementary Note 6). The obtained spiral phase pattern confirms the ability of the phase-engineered nonlinear crystal to generate beams carrying OAM. Indeed, this distribution is the superposition of the Gaussian profile inherited from the pump and the expected azimuthal profile\cite{SrinivasReview}.  

When the imaging system's object plane corresponds to the one of the sample, the four squares are imaged. Figure \ref{fig:Fig3}b shows the spectrally-filtered SH images of the crystal-engineered 3R-MoS$_2$ sample at different sample focal positions, using a pump fluence of $\SI{200}{\micro J/cm^2}$ and an integration time on the CMOS camera of $\SI{99}{ms}$. Upon increasing the objective's distance from the sample, the SH intensity pattern assumes a vortex-like shape, carrying the original square pattern along with the small gaps in-between the individual 3R-MoS$_2$ square slabs, as seen in Fig. \ref{fig:Fig3}b. 

The simulated and measured SH phase profiles for both RCP and LCP excitation are reported in Fig. \ref{fig:Fig3}c and Fig. \ref{fig:Fig3}d, respectively. These images are acquired at a distance of $\SI{0.67}{mm}$ from the object plane, and the phases are obtained subtracting the individual object and reference images to the interference pattern so that the low frequency component of the intensity is filtered out (see Supplementary Note 4 for further details). As expected, the same behaviour of the \SI{46}{nm}-thick sample used in this work is observed on other fabricated 3R-MoS$_2$ samples with identical lateral geometry and crystal engineering design, but different exemplary thicknesses of \SI{48}{nm}, \SI{91}{nm} and \SI{188}{nm}, used as a backup for reproducibility experiments (see Supplementary Note 7).\\

\begin{figure}[H]
    \centering
    \includegraphics[width=.9\linewidth]{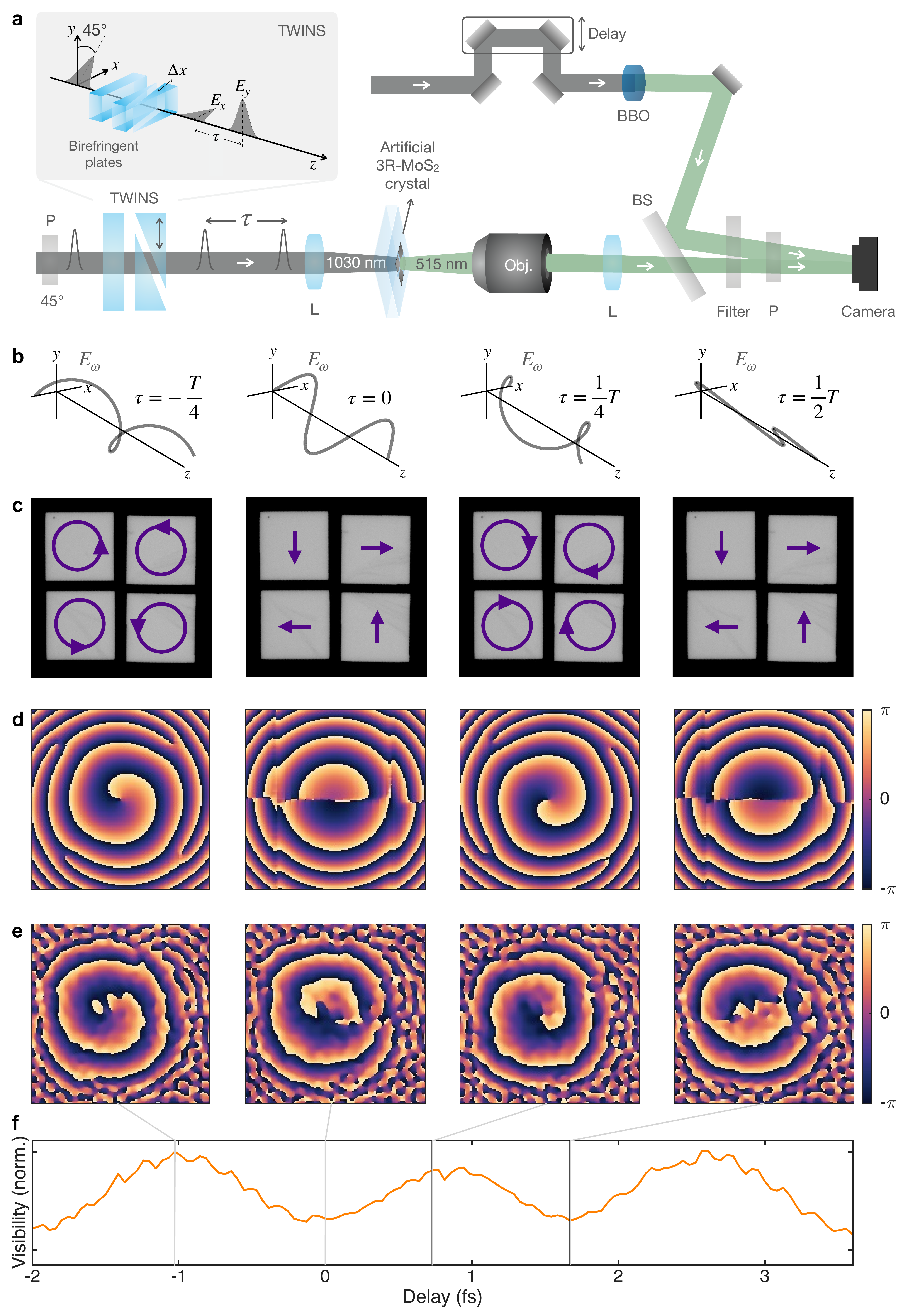}
    \caption{\textbf{Ultrafast control of nonlinear structured light vortexes in crystal-engineered 3R-MoS$_2$. a} Experimental setup. The inset shows the working principle of the TWINS interferometer. \textbf{b} Schematics of the resulting pump field after the TWINS interferometer sketched every $\pi/2$ phase delay. \textbf{c} Schematics of the emitted SH light polarisation in the four different quadrants of the phase-engineered 3R-MoS$_2$ sample in the four polarization configurations. \textbf{d} Simulated phase images of the SH vortex at the corresponding delay between orthogonal excitation fields. \textbf{e} Experimental phase obtained with off-axis digital holography. \textbf{f} Visibility of the interference fringes allowing for the phase reconstruction. The four vertical lines correspond to the four delays at which the above panels have been measured, in the correct order.}  
    \label{fig:Fig4}
\end{figure}

\noindent Finally, exploiting the C$_{3v}$ point-group symmetry of 3R-MoS$_2$, we demonstrate ultrafast all-optical switching of OAM in our crystal-engineered 3R-MoS$_2$ platform. In particular, we are able to switch between a HG and a vortex SH beam, and between vortex beams of opposite topological charge ($l = -1$ and $l = +1$), by controlling the relative delay of a collinear excitation pulse pair with sub-fs resolution.

In our phase-engineered 3R-MoS$_2$ sample, we exploit the linear relationship between crystal orientation and emitted SH phase for the case of CP excitation (Eq. \ref{eq:Pr}-\ref{eq:Pl}) to impress an azimuthal phase term. Note that the same result cannot be obtained with LP excitation because, as Eq. \ref{eq:Px}-\ref{eq:Py} show, the crystal orientation results in a modulation of the electric field's amplitude rather than its phase. The only phase change in this configuration is a $\pi$ shift between the upper and lower section of the structure, which is associated with an HG like phase pattern with a phase singularity along the horizontal or vertical axis (HG$_{01}$ or HG$_{10}$ like). Therefore, with the phase-engineered sample presented in this work, it is possible to switch between HG-like and vortex beams with topological charge $l=\pm1$ by changing the polarisation state of the pump between linear and circular with sub-fs temporal resolution.

To this end, we insert the TWINS common-path birefringent interferometer \cite{TWINS} before the focusing lens, as sketched in Fig. \ref{fig:Fig4}a. The interferometer is composed of two birefringent wedges whose ordinary and extraordinary axes are oriented in such a way that a $\SI{\pm45}{\degree}$ linearly-polarized input field is split into two orthogonal replicas $E_x$ and $E_y$. A relative delay $\tau$ is introduced by transverse translation of one of the birefringent wedges, as shown in Fig. \ref{fig:Fig4}a (further details are provided in the Methods section). By controlling $\tau$ with sub-optical-cycle accuracy, we switch between circular and linear FF fields by changing the relative phase between $E_x$ and $E_y$. 

Figure \ref{fig:Fig4}b schematises the overall pump polarisation for the four different delays. When the relative delay between the two orthogonal components is $\tau=0$ or $\tau=T/2$, \textit{i.e.}, half of the pump optical cycle $T$, the two fields sum up to a $\SI{\pm45}{\degree}$ linearly-polarized resultant. Instead, when the relative delay is $\tau=\pm T/4$ the two fields have a $\mp\frac{\pi}{2}$ phase difference which results in a circularly polarised field. 

These four different polarisation states, \textit{i.e.}, RCP, LP at -\SI{45}{\degree}, LCP and LP at \SI{45}{\degree}, excite the phase-engineered 3R-MoS$_2$ artificial crystal and generate a SH field  whose polarization is represented schematically in Fig. \ref{fig:Fig4}c for each of the four different quadrants of the sample. First and third panels display RCP and LCP excitations, respectively, with a $\pi/2$ phase shift between adjacent squares, as expected from the $3\theta$ phase acquired over the relative orientation $\theta=\SI{30}{\degree}$ between squares, as by design. Second and fourth panels display the linear case, where upper left and lower right squares have vertical polarisation with opposite phase. The remaining two quadrants generate horizontal polarisation and are also offset by a phase of $\pi$. Crucially, upper and lower halves are individually in phase, generating a HG-like field. 

Figure \ref{fig:Fig4}d shows the simulated SH field phases in the four excitation conditions corresponding to the panel above, with clear distinction between vortex ($\tau=-T/4$ and $\tau=\frac{T}{4}$) and HG beam ($\tau=0$ and $\tau=T/2$). In Fig. \ref{fig:Fig4}e we report the experimental phase obtained with the same excitation power and camera setting as the measurements shown in Fig. \ref{fig:Fig3}c, averaging over 3 images. These phase maps show very good agreement with the simulation and a clear distinction between the vortexes and non HG beams (for quantitative comparison see Supplementary Note 8). 

Figure \ref{fig:Fig4}f shows the visibility of the interference fringes, measured as the mean absolute value of the high frequency, phase-carrying components of the interference image's Fourier Transform. These components are the ones being isolated for the phase measurements, and their magnitude is a direct indication of the interference visibility. The signal obtained is periodic with the maxima and minima corresponding to the vortexes and HG beams respectively, hence with a period which is twice the optical cycle. The vertical lines in the graph indicate the interferometer position in which the phases in Fig. \ref{fig:Fig4}e have been captured, in order of appearance. These results demonstrate all-optical switching between a Gaussian and a vortex SH beam, and between vortex beams of opposite topological charge ($l = -1$ and $l = +1$), with sub-optical-cycle accuracy.\\ 


In conclusion, we have demonstrated all-optical switchable nonlinear structured-light generation in an artificial 46-nm-thick 3R-MoS$_2$ crystal. By combining tensor-driven geometric phase with spatial control of the in-plane crystal orientation, we generated structured SH fields and achieved all-optical switching between HG-like modes and vortex beams with opposite topological charges, without cascaded optical elements.

The crystal-engineering strategy introduced here is inherently scalable to more complex orientation patterns and could enable arbitrary nonlinear phase profiles, including higher-order vortex beams, multiplexed structured-light generation and multifunctional nonlinear wavefront shaping within a monolithic platform. Although the geometric-phase mechanism underlying our approach is intrinsically broadband, incorporating optically resonant nanostructures could provide additional spectral selectivity and polarization control, while strengthening light–matter interactions and increasing conversion efficiency.

Combined with the versatility of van der Waals assembly and recent advances in nonlinear metasurfaces, our results position crystal-engineered layered materials as a promising platform for compact, ultrafast and all-optically-reconfigurable nonlinear structured-light sources in integrated classical and quantum photonics.

\newpage

\section*{Methods}\label{methods}
\subsubsection*{Common-path birefringent interferometer}
TWINS consists of two plates of a birefringent material, with thicknesses $L_a$ and $L_b$ respectively, and optical axes rotated by 90°. For an incident optical waveform with polarization rotated by 45° with respect to the ordinary and extra-ordinary axes, the output waveform splits into two replicas with perpendicular polarizations and delay $\tau$ proportional to $L_a-L_b$.
To continuously change the delay, one of the plates is cut into a pair of wedges, one of which is transversely translated. Depending on the relative delay between the two orthogonal electric field components, the polarisation can be linear ($\pm45$°), circular (left or right), or elliptical (the general combination of the two). The minimum delay $\Delta \tau$ is obtained from $\Delta\tau=\Delta n \space \Delta x \space sin(\alpha)/c$, with $\Delta n$ the difference between the ordinary and extraordinary refractive indexes ($\sim 0.1$ in $\alpha$-BBO at 1030 nm), $\Delta x $ the minimum spatial increment of the wedge's position, $\alpha$ the wedge angle, and $c$ the speed of light. In the current configuration, the minimum delay is $\sim\SI{0.02}{fs}$. 

\subsubsection*{Off-axis digital holography}
Digital holography is an imaging technique capable of reconstructing the spatial profile of the electric field's phase. The SHG vortex ("object" field) is made to interfere with a Gaussian beam ("reference" beam) at the same wavelength generated from a 3 mm thick BBO crystal excited by the same laser source. The object and reference beams are recombined with a 50:50 beam splitter and their polarisation is projected onto the same axis with a linear polariser. The interference fringes resulting from the off-axis superposition of the two wavefronts can be seen as  high spatial frequencies in the 2D Fourier transform of the image (k-space, see Supplementary Note 4) that carry the phase information of the object and reference fields. By isolating one of the two components, shifting it to zero frequency (DC component), and transforming this complex-valued filtered spectrum back to real space, the phase of the image can be reconstructed by selecting the argument of the complex-valued image. By controlling the angle between the object and reference beams the separation between the DC (zero order) component and the sidebands in k-space can be increased, reducing the crosstalk between the two and isolating the phase-carrying component. The polariser after the recombining beam slitter is set at $\SI{45}{\degree}$ so that all four quadrant of the sample are imaged simultaneously with the same brightness. 

\backmatter

\newpage


\bmhead{Acknowledgements}
We thank Dr. Paolo Franceschini for the useful discussions.\\
\noindent C.T. acknowledges the European Union’s Horizon Europe research and innovation programme under the Marie Skłodowska-Curie PIONEER HORIZON-MSCA-2021-PF-GF grant agreement No 101066108. C.T. also acknowledges the Optica Foundation for supporting this research through the 2024 Optica Foundation Challenge. P.V. and C.T. acknowledge Agenzia per la cybersicurezza nazionale under the programme for promotion of XLI cycle PhD research in cybersecurity – 132775\_167\_PolMI. The views expressed are those of the authors and do not represent the funding institution. C.T. and G.C. acknowledge funding from the European Union-NextGenerationEU under the National Quantum Science and Technology Institute (NQSTI) Grant No. PE00000023-q-ANTHEM-CUP H43C22000870001. Z.H.P. and P.J.S. acknowledge support under NSF-AFRL program award NSF ECCS-2529062.

\bmhead{Author contributions}

L.C. and C.T. conceived the idea. P.V. designed the experiment, built the setup, performed the nonlinear measurements and analysed the data. L.C. performed the theoretical calculations. M.R., Z.H.P., V.Q.-C., X.-Y.Z. and P.J.S. prepared the sample. M.R., B.S. and B.U. characterised the sample. F.G. and G.C. contributed to the ultrafast measurements. P.V., L.C. and C.T wrote the manuscript with inputs from all co-authors.

\bibliography{sn-bibliography}


\end{document}